\documentclass[12pt]{article}
\usepackage{amstex}
\usepackage{amssymb}
\begin{document}
\newcommand{\half}{\frac{1}{2}}
\newcommand\sech{{\rm sech}}
\newcommand{\IZ}[1]{\bar{#1}}
\newcommand{\Deta}{\eta^{\dagger}}
\newcommand\Bpsi{\boldsymbol{\psi}}
\title{
\begin{flushright}
  \small UMP-96/96
\end{flushright}
\vskip1.0cm
\large\bf
Exact Solution to the Moment Problem for the XY Chain}
\author
{\large N.S. Witte\footnote{E-mail: {\tt nsw\@physics.unimelb.edu.au}}\\
{\it Research Centre for High Energy Physics,}\\
{\it School of Physics, University of Melbourne,}\\
{\it Parkville, Victoria 3052, AUSTRALIA.} }
\maketitle
\begin{abstract}
We present the exact solution to the moment problem for the spin-$1/2$
isotropic antiferromagnetic XY chain with explicit forms for the 
moments with respect to the N\'eel state, the cumulant generating function,
and the Resolvent Operator. We verify the correctness of the Horn-Weinstein
Theorems, but the analytic structure of the generating function 
$ \langle e^{-tH} \rangle $ in the complex $t$-plane is quite different 
from that assumed by the $t$-Expansion and the Connected Moments 
Expansion due to the vanishing gap.
This function has a finite radius of convergence about $ t=0 $, 
and for large $ t $ has a leading descending algebraic series 
$ E(t)-E_0 \sim At^{-2} $. The Resolvent has a branch cut and 
essential singularity near the ground state energy of the form
$ G(s)/s \sim B|s+1|^{-3/4}\exp(C|s+1|^{1/2}) $.
Consequently extrapolation strategies based on these assumptions
are flawed and in practise we find that the CMX methods are
pathological and cannot be applied, while numerical evidence for
two of the $t$-expansion methods indicates a clear asymptotic
convergence behaviour with truncation order.
\end{abstract}
\eject
\section{Motivation}
Some novel non-perturbative formalisms based on a moment approach
have been proposed recently,
the $t$-Expansion\cite{texp-HW}
and the Connected Moments Expansion (CMX)\cite{cmx-1st-C},
and applied in a number of areas ranging from 
quantum chemistry\cite{cmx-1st-C,cmx-chem-C,cmx-CKSP,cmx-YI,cmx-vmc-YI}, 
lattice gauge theories
\cite{texp-H,texp-HS,texp-HK,texp-M,texp-MSB,texp-HL,texp-vdDH} to 
condensed matter problems\cite{cmx-MBM,cmx-MM,cmx-MPM,cmx-afh-LL,cmx-hm-LL}. 
These formalisms take their starting point from
a finite sequence of low order connected Hamiltonian moments, or
cumulants $\{ \nu_n : n=1 \ldots \} $,
and appeal to ostensibly exact theorems for ground
state properties in order to extract appropriate averages.
In all these applications, the sequence of cumulants was finite
and relatively small, and the effects of truncation and the rate of
convergence was completely unknown. 
This issue has particular practical importance because in many of the
applications the convergence and final estimates were rather
poor considering the effort expended in calculating a large
number of moments. 
As rough rule of thumb, the initial and low 
order estimates were very good and the situation looked promising
but when taken to higher order 
there was not a significant improvement, and in some cases the
convergent behaviour seemed to be arrested\cite{cmx-afh-LL,cmx-LW,cmx-hm-LL}.
The expectation of an accurate, perhaps arbitrarily and controllably
accurate, method has to date seemed to be unfounded.
The source of these problems had not been properly identified nor
any remedy proposed.
Here we give the first example of an exact solution to a moment 
problem which can answer some of these questions. 

Furthermore it was not known whether
these techniques were capable of yielding exact results in any case.
This is a natural expectation since any formalism, 
applicable to general problems, must have a set of simple, 
but nontrivial solvable cases.
There was another model\cite{cmx-hom-C}, 
of quadratically coupled harmonic oscillators, for which the exact
ground state wavefunction and density matrices are known but the
moment problem was not exactly solved in that a general cumulant
was not given.
We present the solution to the moment problem for the isotropic,
one-dimensional antiferromagnetic spin $\half$ XY Model.
Our solution also demonstrates an alternative notion of solvability, 
from those concepts and methods usually associated with the 
known solvable models.

In the $t$-Expansion, one constructs the cumulant generating function
from its taylor series expansion about $ t=0 $,
\begin{equation}
   \langle e^{tH} \rangle = 
   \exp\left\{ \sum^{\infty}_{n=1} \nu_n {t^n \over n!} \right\} 
   \equiv e^{F(t)} \ ,
\label{def-F}
\end{equation}
where all moments are calculated with respect to some trial state,
with the proviso that the overlap with exact ground state is
nonzero.
Then the Horn-Weinstein Theorem\cite{texp-HW} states that the ground 
state energy and first excited state gap (within the same sector as
the ground state) are given by the following 
limits
\begin{equation}
 \begin{split}
  E_0     & = \lim_{t \to \infty} E(t) \\
  E_1-E_0 & = \lim_{t \to \infty} 
          -{ d^2E(t)/dt^2 \over dE(t)/dt} \ ,
\end{split}
\label{HW-thm}
\end{equation}
where $ E(t) \equiv F'(-t) $.
Other ground state averages can be found as well by computing the
appropriate generalised connected moments\cite{texp-HW} and 
performing the same extrapolation. 
This theorem was ``proved'' in Ref.~(\cite{texp-HW}) by assuming
a simple discrete point spectrum for the system, even in the
thermodynamic limit, and these kinds of assumptions continue to
be commonly used.
The question immediately arises of how this extrapolation can be
carried out and several methods\cite{texp-S} are employed. 
These are 
\begin{description}
 \item[\rm Pad\'e Analysis:]\mbox{}\\
  $ E(t) $ is assumed to be representable by a $ [L/M] $ Pad\'e
  approximant with $ L=M $ i.e. it has only separable poles and
  zeros of finite order in any finite region of the complex
  $t$-plane, and the extrapolation $ t \to \infty $ made directly,
 
 \item[\rm D-Pad\'e Analysis:]\mbox{}\\
  $ dE(t)/dt $ is assumed to be representable by a $ [L/M] $ Pad\'e
  approximant with $ M \geq L+2 $, on the basis of the same 
  considerations as above, and the ground state energy is found by
  numerically integrating this
  \begin{equation}
    E(\infty)-E(0) =
    \int^{\infty}_{0} ds E'(s) \ ,
  \label{analytic-D}
  \end{equation}
 
 \item[\rm Laplace Method/Resolvent Analysis:]\mbox{}\\
  The Resolvent 
  \begin{equation}
    G(u)/u \equiv \langle {1\over 1+uH} \rangle \ ,
  \label{def-G}
  \end{equation}
  is assumed to have a pole structure of the form
  \begin{equation}
    G(u)/u \sim 
    {b_0 \over 1+uE_0} + {b_1 \over 1+uE_1} + \ldots \ ,
  \label{analytic-G}
  \end{equation}
  so that a $ [L/M] $ Pad\'e analysis with $ M=L+1 $ can be made
  and the pole found,

  \item[\rm Inversion Analysis:]\mbox{}\\
  One inverts the $ E(t) $ function, forms $ dt(E)/dE $ and makes 
  a Pad\'e analysis of this. The ground state energy is the location
  of a pole in this derivative, from
  \begin{equation}
     {dt \over dE} \sim {C \over E-E_0} \ ,
  \label{analytic-t}
  \end{equation}

  \item[\rm ``$ E $ of $ F $''/Partition Function:]\mbox{}\\
  The function $ \delta(t) = 1-\langle e^{-tH} \rangle $ is taken
  as the independent variable instead of $ t $.
  One inverts the $ \delta(t) $ function and makes a diagonal Pad\'e
  analysis of $ E(\delta) $ and extracts the ground state energy
  from
  \begin{equation}
     E_0 = \lim_{\delta \to 1^-} 
         {1 \over (1\!-\!\delta)
         {\displaystyle dt \over \displaystyle d\delta}} \ .
  \label{extrap-delta}
  \end{equation}
\end{description}

The Connected Moments Expansion, of which there are many variants, 
is derived by assuming 
\begin{equation}
   E(t) \sim E_0 + \sum^{n}_{j=1} A_j e^{-b_j t} \ .
\label{cmx-Et}
\end{equation}
The variants give the following explicit forms for the ground state
energy
\begin{description}

  \item[\rm CMX-HW:]\mbox{}\\
  The first method\cite{cmx-1st-C} gives the ground state energy up
  the $n$th order as
  \begin{equation}
     E_0 = \nu_1-\sum^{n}_{i=1} 
     { S^2_{i,2} \over \prod^{i}_{j=1} S_{j,3} } \ ,
  \label{cmxhw-gse}
  \end{equation}
  where the $ S $-variables are recursively generated by
  \begin{equation}
  \begin{split}
     S_{1,k} & = \nu_k \\
     S_{i+1,k} & = S_{i,k}S_{i,k+2}-S^2_{i,k+1} \ ,
  \end{split}
  \label{cmxhw-recur}
  \end{equation}

  \item[\rm CMX-SD:]\mbox{}\\
  Another method\cite{cmx-hom-C} gives the ground state energy up
  the $n$th order as
  \begin{equation}
     E_0 = \omega_{1,n} \ ,
  \label{cmxsd-gse}
  \end{equation}
  where the $ \omega $-variables are recursively generated by
  \begin{equation}
  \begin{split}
     \omega_{j,1} & = \nu_j \\
     \omega_{j,k+1} & = 
     \omega_{j,k}-2{ \omega_{j+1,k}\omega_{2,k} \over \omega_{3,k} }
     +\left({ \omega_{2,k} \over \omega_{3,k} } \right)^2\omega_{j+2,k} \ ,
  \end{split}
  \label{cmxsd-recur}
  \end{equation}

  \item[\rm CMX-LT:]\mbox{}\\
  This method described in Ref.~\cite{cmx-hom-C,cmx-K} gives
  \begin{equation}
     E_0 = \nu_1
     - (\nu_2,\nu_3, \ldots ,\nu_n) \left(
       \begin{array}{cccc}
        \nu_3      &  \nu_4      & \cdots  &  \nu_{n+1} \\
        \nu_4      &  \nu_5      & \cdots  &  \nu_{n+2} \\
        \vdots     &  \vdots     & \ddots  &  \vdots    \\
        \nu_{n+1}  &  \nu_{n+2}  & \cdots  &  \nu_{2n-1}
       \end{array}                  \right)^{-1}
                                    \left(
       \begin{array}{c}
        \nu_2   \\
        \nu_3   \\
        \vdots  \\
        \nu_{n} 
       \end{array}                  \right) \ .
  \label{cmxlt-gse}
  \end{equation}
\end{description}
There is also another variant, called the AMX\cite{cmx-MZM}, which
is very similar to the CMX-HW method.
The CMX, in addition to its convergence difficulties, often presents
another, although less serious problem. Usually the cumulants are
nonlinear functions of some coupling constant, and the resulting
ground state energy estimate above truncated at some order above would 
be rational function of the constant. 
In these problems\cite{cmx-U,cmx-MMPZ} an unphysical
pole would appear for real coupling values. 
This pole was found to be exactly cancelled\cite{cmx-MZM} if one 
included the next order, but a next order is not always available.

\section{The Model}

We follow the conventions and notations of Ref.~\cite{xy-LSM} as closely 
as possible in what follows. For the $ N $ site chain, with $ N $ even
but not a multiple of 4, the Hamiltonian is taken to be
\begin{equation}
 H = \sum^{N}_{i=1} \left[ S^x_iS^x_{i+1}+S^y_iS^y_{i+1} \right] \ ,
\label{xy-ham}
\end{equation}
where the labels $ i,j,\dots $ index the sites.
Following the standard treatments a transformation from spin operators
to a mixed Fermi-Bose operator algebra is performed
\begin{equation}
  a^{\dagger}_i, a_i \equiv S^x_i \pm iS^y_i \ ,
\label{bose-spin}
\label{BF-xfm}
\end{equation}
and then a Wigner-Jordan transformation to a pure Fermionic representation
\begin{equation}
\begin{split}
  c_i & \equiv 
      e^{\pi i \sum^{i-1}_{j=1} a^{\dagger}_j a_j} a_i \\
  c^{\dagger}_i & \equiv 
      a^{\dagger}_i e^{-\pi i \sum^{i-1}_{j=1} a^{\dagger}_j a_j} \ ,
\end{split} 
\label{wj-xfm}
\end{equation}
so that the Hamiltonian is now
\begin{equation}
 H = \half \sum^{N}_{i=1} 
     \left[ c^{\dagger}_{i}c_{i+1}+c^{\dagger}_{i+1}c_{i} \right] \ .
\label{fermi-ham}
\end{equation}
In doing so we have neglected the boundary terms arising from the exact
mapping of the spin or a-cyclic problem, and will only consider the
c-cyclic problem $ c_{N+1} = c_1 $. We take this course because we don't
want to complicate matters unnecessarily and are only interested in
the thermodynamic limits.

Proceeding with the diagonalisation of this quadratic form via the
canonical transformation
\begin{equation}
\begin{split}
  \eta_p
  & = \sum_i \left[  {\phi_{pi}\!+\!\psi_{pi} \over 2} c_i
                   + {\phi_{pi}\!-\!\psi_{pi} \over 2} c^{\dagger}_i
             \right]
  \\
  \Deta_p
  & = \sum_i \left[  {\phi_{pi}\!+\!\psi_{pi} \over 2} c^{\dagger}_i
                   + {\phi_{pi}\!-\!\psi_{pi} \over 2} c_i
             \right] \ ,
\end{split} 
\label{diag-xfm}
\end{equation}
with transformation coefficients
\begin{alignat}{2}
  \phi_{pi}
  & = \sqrt{2\over N} \sin pi
  & \qquad p & >    0 \notag \\
  & = \sqrt{2\over N} \cos pi
  & \qquad p & \leq 0 \notag \\
  \psi_{pi}
  & = {\cos p\over \Lambda_p} \phi_{pi}
  & \qquad   & \ ,
\label{coeff-xfm}
\end{alignat}
where the quasiparticle energy $ \Lambda_p = |\cos p\,| $ and the
momentum values, labelled by $ p,q,\ldots $ are
\begin{equation}
   p = {2\pi m\over N}  \qquad
   m = -N/2, \ldots ,0,1, \ldots , N/2-1 \ .
\label{momenta}
\end{equation}
The Hamiltonian then becomes
\begin{equation}
  H = \sum_p \Lambda_p \Deta_p\eta_p 
            - \half \sum_p \Lambda_p  \ .
\label{ham-diag}
\end{equation}
The transformation coefficients are orthonormal in the sense
\begin{equation}
\begin{split}
 \phi_p \cdot \phi_{p'}
 & = \delta_{pp'} \\
 \psi_p \cdot \psi_{p'}
 & = \delta_{pp'} \\
 \phi_p \cdot \psi_{p'}
 & = 0 \ .
\end{split} 
\label{coeff-orth}
\end{equation}

In all the moment formalisms a choice has to be made as what trial
state the moments of the Hamiltonian are computed with respect to
and it is this choice which is the crucial physical one. The only
general requirement is that, for ground state properties at least,
the trial state should have a non-zero overlap with the true
ground state. In all the previous and approximate treatments of the
the XY chain, and most of the other antiferromagnetic spin models,
the trial state was taken to be the classical N\'eel state. 
This state is in fact, as we shall quantify, as far from the true
ground state as one could possibly be but its simplicity allows
for ease of moment computation, certainly compared to alternatives.
So as a ``natural'' choice we proceed by only considering this 
choice.

In the spin picture the N\'eel state $ |N\rangle $ will be represented by
\begin{equation}
  | \uparrow \downarrow \uparrow \downarrow \ldots 
    \uparrow \downarrow \uparrow \downarrow \rangle \ ,
\label{s-neel}
\end{equation}
and in the fermionic $c-$picture by
\begin{equation}
  | 1 0 1 0 \ldots 1 0 1 0 \rangle \ .
\label{c-neel}
\end{equation}
The defining equations for the N\'eel state are
\begin{equation}
  \begin{aligned}
    c^{\dagger}_{2i+1} |N \rangle & = 0 \\
    c_{2i} |N \rangle             & = 0
  \end{aligned}
  \Biggr\} \quad \forall i \,
\label{neel-defn}
\end{equation}
and using the transformation coefficients, Eq.~(\ref{coeff-xfm}), and 
the spectral points, Eq.({\ref{momenta}),
these can be shown to be equivalent to the following equations in 
terms of the quasiparticle operators
\begin{equation}
  \begin{aligned}
    (\eta_{p}-\Deta_{\bar{p}}) |N \rangle & = 0 \\
    (\Deta_{p}+\eta_{\bar{p}}) |N \rangle  & = 0
  \end{aligned}
  \Biggr\} \quad \forall p \in Z \ .
\label{neel-quasi}
\end{equation}
Here the spectrum of allowed momentum values have been divided up into 
the set
\begin{equation}
   Z \equiv [-\pi, -\pi/2) \cup (0,\pi/2) \ ,
\label{z-defn}
\end{equation}
and its complement $ \bar{Z} $, and there is a bijection from $ Z $ to
$ \bar{Z} $ defined by
\begin{alignat}{2}
  \bar{q}
  & \equiv -q-\pi
  & \qquad q & \leq 0 \notag \\
  & \equiv -q+\pi
  & \qquad q & >    0 \ .
\label{z-map}
\end{alignat}

If we denote the vacuum state for the quasiparticles by 
$ |0\rangle $ then the N\'eel state can be expressed in terms of 
the quasiparticles by
\begin{multline}
  |N\rangle = 2^{-N/4} \Biggl\{
   1 + \sum_{q \in Z} \Deta_q\Deta_{\IZ{q}}
   + \sum_{q_1<q_2 \in Z} 
         \Deta_{q_1}\Deta_{\IZ{q}_1}\Deta_{q_2}\Deta_{\IZ{q}_2}
                       \\
                       \ldots +
         \Deta_{q_1}\Deta_{\IZ{q}_1}\Deta_{q_2}\Deta_{\IZ{q}_2}\ldots 
         \Deta_{q_{N/2}}\Deta_{\IZ{q}_{N/2}}
                       \Biggr\} |0\rangle \ ,
\label{neelrep-1}
\end{multline}
and because of the anticommutation relations for the quasiparticle operators
\begin{equation}
\begin{split}
  |N\rangle
  & = 2^{-N/4} \exp \left( \sum_{q\in Z} \Deta_q\Deta_{\IZ{q}} \right)
               |0\rangle
  \\
  & = \prod_{q\in Z} {1+\Deta_q\Deta_{\IZ{q}} \over \sqrt{2}}
               |0\rangle \ .
\end{split} \label{Neelrep}
\end{equation}
One can easily show that this state satisfies the defining 
Eq.~(\ref{neel-quasi}).
Basically the N\'eel state is created from the quasiparticle vacuum state
by pairs of excitations whose momenta sum to $ \pm\pi $.

We will have need of some spectral sums defined on $ Z $ and these are
\begin{equation}
\begin{split}
 \sum_{q\in Z} \Lambda^{2n}_q
 & = { (2n)! \over 2^{2n+1}(n!)^2 } N
 \\
 \sum_{q\in Z} \Lambda_q
 & = { 1\over \sin(\pi/N) } \ .
\end{split}
\label{q-sums}
\end{equation}

\section{Generating Function}

We now show how to find the exact form for the generating function
\begin{equation}
   \langle e^{tH} \rangle \equiv 
   \langle N| e^{tH} |N\rangle
   = \sum^{\infty}_{n=0} {\mu_n t^n\over n!} \ ,
\label{gf-exp}
\end{equation}
where the moments $ \mu_n \equiv \langle H^n \rangle $.
We will establish this result for a finite system of size $ N $
although ultimately the thermodynamic limit will be taken.
It is straightforward to show a direct expression for a general moment
\begin{multline}
 \mu_n = 2^{-N/2} \Biggl\{
                  E^n_0 + \sum_{q\in Z} (E_0+2\Lambda_q)^n
       + {1\over 2!}\sum_{q_1 \neq q_2\in Z} 
                 (E_0+2\Lambda_{q_1}+2\Lambda_{q_2})^n
 \\
       \ldots +
       (E_0+2\Lambda_{q_1}+ \ldots +2\Lambda_{q_{N/2}})^n
                  \Biggr\} \ .
\label{moment}
\end{multline}
in its symmetrical form. 
Then our initial form for the generating function is
\begin{multline}
 \langle e^{tH} \rangle
       = 2^{-N/2} e^{E_0 t} \Biggl\{
                  1 + \sum_{q\in Z} e^{2\Lambda_q t}
       + {1\over 2!}\sum_{q_1 \neq q_2\in Z} 
                 e^{2t(\Lambda_{q_1}+\Lambda_{q_2})}
 \\
       \ldots +
       e^{2t(\Lambda_{q_1}+ \ldots +\Lambda_{q_{N/2}})}
                  \Biggr\} \ .
\label{gf-start}
\end{multline}
For simplicity let us define the following sum
\begin{equation}
   I_n \equiv \sum_{q\in Z} e^{2nt\Lambda_q} \ ,
\label{q-expsum}
\end{equation}
and denote the general $p$-th term of the Eq.~(\ref{gf-start})
by $ g_p/p! $ after expanding it. 
This would now be
\begin{equation}
   {g_p \over p!} =
   {1\over p!} \sum_{q_1 \neq \ldots \neq q_p \in Z}
                e^{2t\Lambda_{q_1}} \ldots e^{2t\Lambda_{q_p}} \ .
\label{avoid-sum}
\end{equation}
Such sums over mutually distinct spectral points are recognisable
as the cycle index of the Symmetric Group $ S_p $.
Consequently we have the partition sum formula
\begin{equation}
  {g_p \over p!} =
  \sum_{ \sum li_l = p } 
  { 1\over i_l!1^{i_l} i_2!2^{i_2} \ldots i_p!p^{i_p} }
  (-)^{p+\sum i_l} I_1^{i_1} \ldots I_p^{i_p} \ ,
\label{cycle-index}
\end{equation}
and find immediately the exponential sum, constituting the generating
function, is
\begin{equation}
   \sum^{\infty}_{p=0} {g_p \over p!} =
   \exp \left\{ \sum^{\infty}_{n=1} {(-)^{n+1} \over n}I_n \right\} \ .
\label{cycle-id}
\end{equation}
Each term $ g_p/p! $ is an elementary symmetric polynomial in the
variables $ z_p \equiv e^{2t\Lambda_p} $ and Eq.~(\ref{cycle-index})
is just the way of re-expressing this in terms of the power sums
$ I_n = \sum_p z^n_p $.

The generating function is then expressible in the following
sequence of closed forms
\begin{equation}
\begin{split}
  \langle e^{tH} \rangle
  & =  2^{-N/2}e^{tE_0}
       \exp\left\{ \half\log \prod_{q}(1+e^{2t\Lambda_q}) \right\}
  \\
  & =  \prod_{q\in Z} \cosh(t\Lambda_q)
  \\
  & =  \prod^{\infty}_{k=0} \prod_{q\in Z}
       \left( 1+ {1\over \pi^2(k\!+\!1/2)^2} t^2\Lambda^2_q \right)
  \\
  & @>>{N \to \infty}>
       \exp\left\{ {N\over 4\pi} 
           \int^{\pi}_{-\pi}\, dq \log\cosh(t\Lambda_q) \right\} \ .
\end{split}
\label{gf-result}
\end{equation}
This leads to the generating function $ F(t) $ associated with the 
$t$-expansion
\begin{equation}
  F(t) = {N \over \pi} 
           \int^{\pi/2}_{0}\, dq \log\cosh(t\Lambda_q) \ ,
\label{ft-result}
\end{equation}
which is identical to the form of the free energy\cite{xy-LSM} for 
this system of noninteracting quasiparticles with zero chemical 
potential, with the identification $ \half\beta \to t $.
The corresponding density $ f(t) $, defined by $ F(t) \equiv Nf(t) $ is,
\begin{equation}
  e^{f(t)} = \prod^{\infty}_{k=0} \left[ \half 
           + \half\sqrt{ 1+{t^2 \over \pi^2(k\!+\!1/2)^2} } \right] \ .
\label{exp-ft}
\end{equation}

This function is also the cumulant generating function, so from the
taylor series expansion about $ t=0 $ we have
\begin{alignat}{2}
  \nu_{2n+1}
  & = 0
  & \qquad n & \ge 0 \notag \\
  \nu_{2n}
  & = {2^{2n}\!-\!1 \over 4n}{(2n)! \over (n!)^2} B_{2n} N
  & \qquad n & \ge 1 \ ,
\label{cumulant}
\end{alignat}
where $ B_{2n} $ are the standard Bernoulli numbers.
These cumulants coincide exactly with all the low order 
ones\cite{pexp-1d-H} computed directly by combinatorial 
enumeration on a computer.

There is also a simpler, more useful expression for the moments,
other than that given by Eq.~(\ref{moment}).
Using the infinite product expansion for the generating function
and expanding it in $ t^2 $, one collects all factors contributing
to $ t^{2n} $ term. By picking $ n $ such factors, all of which
must be distinct,
\begin{equation}
   {\mu_{2n} \over (2n)!} = 
   {1\over n!}\left({4\over \pi^2}\right)^n
   \sum_{(k_1,q_1) \neq \ldots \neq (k_n,q_n)}
   {\Lambda^2_{q_1} \over (2k_1\!+\!1)^2} \ldots
   {\Lambda^2_{q_n} \over (2k_n\!+\!1)^2} \ ,
\label{moment-kq-sum}
\end{equation}
and using
\begin{equation}
   \sum^{\infty}_{k=0} {1\over (2k\!+\!1)^{2i}}
   \sum_{q\in Z} \Lambda^{2i}_q
  = { \pi^{2i} (2^{2i}\!-\!1) \over 2^{2i+2}(i!)^2 }
    |B_{2i}| N \ ,
\label{kq-id}
\end{equation}
one can apply the cycle index formula to arrive at
\begin{equation}
  \mu_{2n} = (2n)!
  \sum_{\sum li_l = n} \prod^{n}_{l=1}
  {1\over i_l!}
  \left( {(-)^{l+1}(2^{2l}\!-\!1) \over 4l(l!)^2} |B_{2l}| N
  \right)^{i_l} \ .
\label{moment-simple}
\end{equation}

\section{Analyticity and Analysis}
 In this part we analyse our results for the generating function 
and its closely related functions. 
The cumulant generating function $ F(t) $ and its derived functions 
have a finite radius of convergence about $ t=0 $ of $ |t| < \pi/2 $,
and which is evident from the infinite product form of 
Eq.~(\ref{exp-ft})
in having branch points at $ t=\pi i(k+1/2) $ for all integer $k$.
This finite radius of convergence about $ t=0 $ will then be a source 
of problems for extrapolation methods, even though the function itself
is analytic in the neighbourhood of $ \Re(t) \ge 0 $.

Focussing on
\begin{equation}
  E(t) = -{N\over \pi} \int^{\pi/2}_0 dq\,
          \Lambda_q \tanh(t\Lambda_q) \ ,
\label{et-result}
\end{equation}
we verify the Horn-Weinstein Theorem is exactly satisfied
\begin{equation}
  \lim_{t \to \infty} E(t) = -{N \over\pi} \ .
\label{et-hw-thm}
\end{equation}
Now the first excited state has an energy $ X_1 = E_0+2\Lambda_q $
for a value of 
\begin{equation}
    q = \pm{\pi \over 2}(1\pm {2\over N}) \ ,
\label{low-quasip}
\end{equation}
so that $ X_1-E_0 \sim 2\pi/N $ and that the gap vanishes like 
$ {\rm O}(1/N) $. The Horn-Weinstein theorem for the gap also
gives zero.
Finally the expression for the overlap\cite{cmx-overlap-C} 
\begin{equation}
   |\langle N | 0 \rangle |^2
  = \exp\left\{ -\int^{\infty}_0 dt [E(t)-E(\infty)] \right\} \ ,
\label{overlap}
\end{equation}
can also be shown to give the correct answer of $ 2^{-N/2} $.
The vanishing of this in the thermodynamic limit is another 
signal of the problems with using the N\'eel state.

However while the specific predictions of the theorem are correct
the basis for these are not, as will be evident in what follows.
We require for this the large $ t $ development of the function
$ E(t) $ and this can be obtained in a number of ways. Using the
Mellin Inversion Theorems one can get the leading series only,
so we adopt a direct expansion of the integrand.
This yields
\begin{multline}
   E(t)-E_0 =
   {N\over 2\pi} \sum^{\infty}_{m=1}
   {\Gamma(m\!-\!1/2) \over m!\Gamma(1/2)} (1\!-\!2^{1-2m}) 
   |B_{2m}| \left({\pi \over t}\right)^{2m}
   \\
   + {2N\over \pi} \sum^{\infty}_{n=1} (-)^ne^{-2nt}
     \sum^{\infty}_{m=0} {(1/2)_m (2m\!+\!1)! \over m!}
     \sum^{2m+1}_{k=0} {1 \over k!}(2nt)^{-2m-2+k} \ .
\label{gf-large-t}
\end{multline}
Several comments are in order here. The above $m$ sums are separately
divergent sums, and consequently these expansions are asymptotic, but
when combined the result is convergent.
The leading terms are algebraic, and the exponential terms are 
subdominant, with a numerical evaluation
\begin{equation}
   E(t)-E_0 \sim {N\over 2\pi}
  \left\{ {1\over 12} \left({\pi \over t}\right)^2 
        + {7\over 960}\left({\pi \over t}\right)^4 + \ldots \right\} \ ,
\label{et-2terms}
\end{equation}
which is at complete variance with common assumptions discussed above. 
This is basically due to the vanishing spectral gaps.
The specific leading exponent of $2$ and not $1$ in the descending series
of $t$ is also significant and different from the Pad\'e analysis
assumptions.
A diagonal P\'ade analysis of the taylor series for $ E(t) $ is not 
possible because the $ \nu_{2n+1} $ are zero, 
and so the $ t \to \infty $ limit either doesn't exist or is trivially zero. 

However a D-P\'ade analysis is possible as [M/M+2] approximants can be
constructed for $ E'(t) $. We display in Figure 1 the error for
the ground state energy, computed using a series of D-P\'ade 
approximants for $ M $ values up to 24. It is clear that while 
there is initial convergence, this eventually ceases and no further
improvement is evident. And the best result is never any better than
$ 10^{-4} $. Furthermore there are some gaps in our data where the 
P\'ade approximant had singularities close to the positive real axis 
which made the numerical integration algorithm unreliable.
There are two sources of error in this analysis, the
first being the error due to the finite P\'ade approximants of the
integrand, and the second due to the numerical integration. We consider
the latter error to be small (much smaller than $ 10^{-4} $) and 
relatively uniform across the series so that most of the variation 
is due to the truncation.

The Inversion Method is also flawed here in that the leading order
factor for $ t(E) $ has a branch cut singularity, not a pole
\begin{equation}
  t(E) \sim \left({\pi\over 24}N\right)^{1/2}
            {1 \over \sqrt{E-E_0}} \ .
\label{invers-t}
\end{equation}
We have done an inversion analysis, using a full range of P\'ade
approximants $ [L/M] $ to Eq.~(\ref{analytic-t}), 
and present the error for the ground state energy in Figure 2 
against $ L $.
There are some points not evident in this Figure, and that is the
root of the denominator polynomial closest to the exact ground
state energy was not necessarily real, nor the one with the lowest
real part. Here it is also clear that there is good initial 
convergence but the improvement ceases and in fact gets worse.
There are some points much closer to the true answer than the rest
but it is also true that there is clustering around this point
and there is no way of uniquely distinguishing this point. Also
going to higher order doesn't lead to any improvement.

The Resolvent can be exactly written down as
\begin{equation}
  G(s)/s =
  \int^{\infty}_0 dx\,
  \exp\left\{ {N\over \pi}
      \left[-x + \int^{\pi/2}_0dq\,\log\cosh(sx\Lambda_q)
      \right] \right\} \ ,
\label{resolvent-result}
\end{equation}
in terms of a scaled inverse energy $ s=uN/\pi $, although not 
explicitly evaluated.
Instead we present a saddle point approximation to this integral
for $ N \to \infty $ which yields
\begin{multline}
  G(s)/s \sim
  \left\{ {N\over 2\pi^2}s^2 
          \int^{\pi/2}_0 dq\, \Lambda^2_q \sech^2(sx_0\Lambda_q)
  \right\}^{-1/2}
  \\
  \times \exp\left\{ {N\over \pi}
      \left[ -x_0 
             +\int^{\pi/2}_0 dq\, \log\cosh(sx_0\Lambda_q) \right]
      \right\} \ ,
\label{res-saddle-int}
\end{multline}
where the saddle point $x_0$ is given by the functional equation
\begin{equation}
   1 = s\int^{\pi/2}_0 dq\, \Lambda_q\tanh(sx_0\Lambda_q) \ .
\label{res-saddle-pt}
\end{equation}
The inverse energy lies in the range $ |s|>1 $ and as $ s \to -1 $
then the point $ x_0 \to \infty $. 
Using the asymptotic form for this integral, 
namely from the large $t$ analysis Eq.~(\ref{et-2terms}), 
one finds in the neighbourhood of the ground state energy
\begin{equation}
  x_0 \sim {\pi \over 2\sqrt{6}}|s+1|^{-1/2} \ ,
\label{approx-sp}
\end{equation}
and that the Resolvent behaves like
\begin{equation}
  G(s)/s \sim
  2^{-N/2} \left({4.6^{3/2}\over \pi^{7/2}}N \right)^{-1/2}
  |s+1|^{-3/4} \exp\left({N \over 2\sqrt{6}}|s+1|^{1/2} \right) \ .
\label{approx-resolvent}
\end{equation}
So that we have a branch cut and essential singularity, and not simple
separated poles.

For the ``$E$ of $F$'' analysis the inversion of the function $ \delta(t) $
for $ \delta \to 1^- $ has the following leading order term
\begin{equation}
   t = {\pi \over 2}\log 2 + {\pi \over N}\log(1-\delta)
       + {\rm O}(1/\log(1-\delta)) \ ,
\label{delta-inv}
\end{equation}
and consequently the energy function $ E(\delta) $ has the leading term
as $ \delta \to 1^- $ (with fixed $ N $) of
\begin{equation}
  E(\delta)-E_0 \sim {N^3 \over 24\pi}
                {1 \over [\log(1\!-\!\delta)]^2} \ ,
\label{delta-e}
\end{equation}
which is not a rational function of $ \delta $.

The Connected Moments Expansions actually exhibit pathological behaviour
in this Model in that for both the CMX-HW and CMX-SD the recursion
can't even get started because $ \nu_3 = 0 $. And for the CMX-LT
when the dimension of the matrix to be inverted is odd then it is exactly
singular, and when it is even the inner product is exactly zero, 
which is again due to $ \nu_3 $ vanishing. This behaviour is 
typical in that the expansions work better when $ \nu_3 $ is large,
say relative to $ \nu^{3/2}_2 $.

\section{Conclusions}
  We have given the exact solution to the moment problem for the
one dimensional isotropic spin-half antiferromagnetic XY Model 
and explicitly found
expressions for the moments, the generating function and the 
Resolvent. With this Model we have shown the Theorems used by
the $t$-Expansion and Connected Moment Expansion are correct but
that the assumptions upon which they are based, are not true. 
In fact we find leading order algebraic decay in $ E(t)-E_0 $ of
second degree and not sums of exponential decay terms, that the
inversion of $ E(t) $ has a leading order branch cut, and that the
Resolvent in the neighbourhood of the ground state energy has
branch cut and essential singularities. 
For this Model we find the CMX methods and some of the $t$-expansion 
extrapolation methods to give pathological results, and the two
$t$-expansion methods that can be applied give clear indications
of having asymptotic convergence behaviour with respect to the 
truncation order.
The analytic properties of this Model are at
complete variance with commonly held assumptions made by the
methods and shows why extrapolation and expansion procedures based
on these assumptions have poor convergence.
However we optimistically predict that revision of the extrapolation
methods, along lines indicated by our results, should dramatically
improve the convergence characteristics of the $t$-Expansion and
CMX-type methods. Although the XY Model is perhaps the most difficult
case to treat with moment methods employing the N\'eel state as the
trial state, being the most extreme quantum limit of the anisotropic
antiferromagnetic Heisenberg Models, a number of features found here
are expected to survive for the isotropic Heisenberg Model. In this
last case the excited state gap still vanishes, as in our current
example.
We also want to emphasize that the general questions of convergence 
in these methods, which are primarily of a physical and not just
a technical mathematical nature, still remain open.

\eject

\bibliographystyle{aip}
\bibliography{xy,texp,cmx,pexp}
\vspace*{30mm}
\begin{center}
\bf Acknowledgements
\end{center}
Part of this work was supported by the Australian Research Council.
The author would like to acknowledge encouragement from and valuable 
discussions with Lloyd C.L. Hollenberg.
\vfill\eject

\listoffigures
\begin{figure}[htb]
 \vskip 20.0cm
 \includegraphics{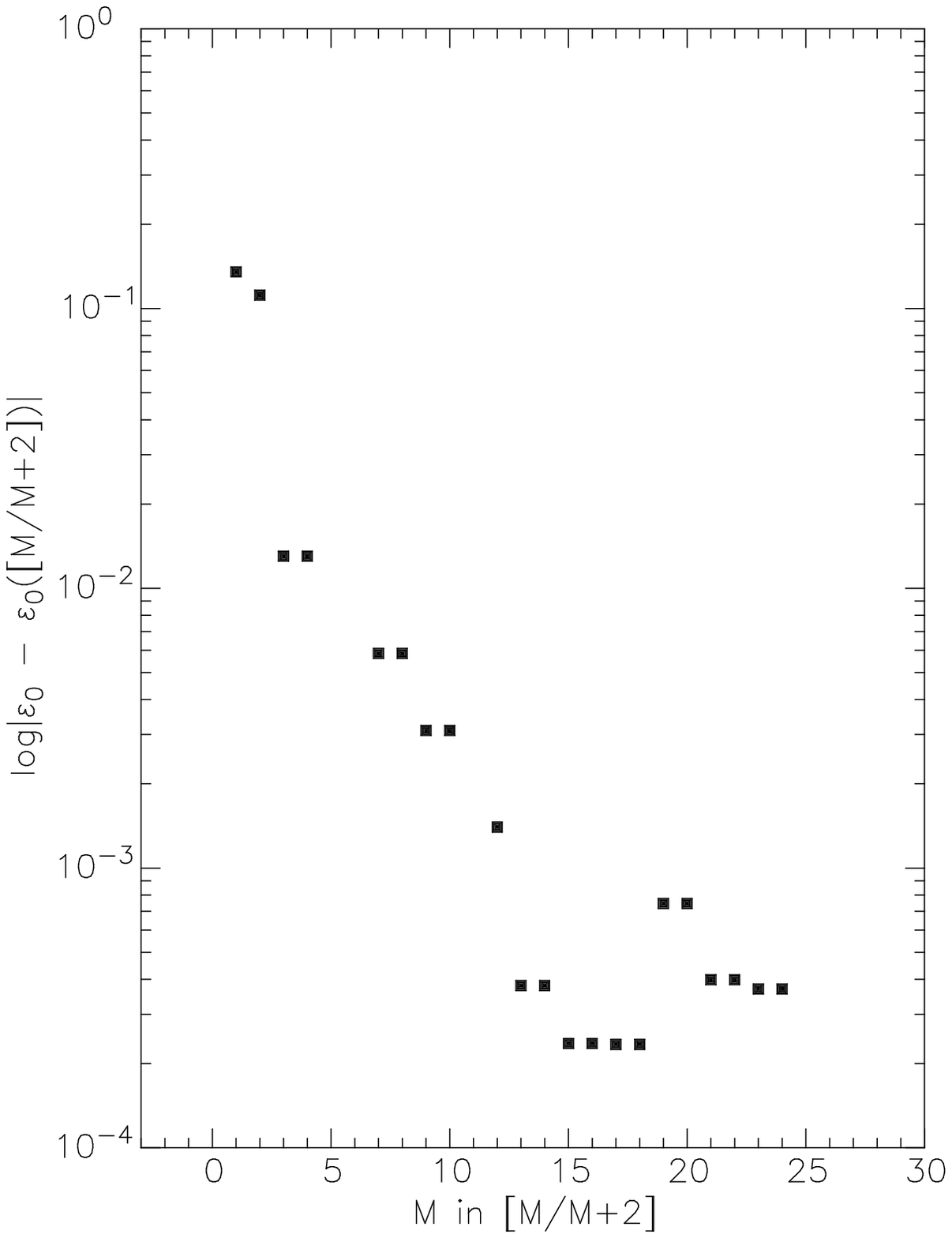}
 \caption
   [The error for the ground state energy computed using a D-P\'ade
    analysis of $ dE(t)/dt $ with a \hbox{$ [M/M+2] $} approximant, 
    versus $ M $.]{}
\end{figure}
\eject

\begin{figure}[htb]
 \vskip 20.0cm
 \includegraphics{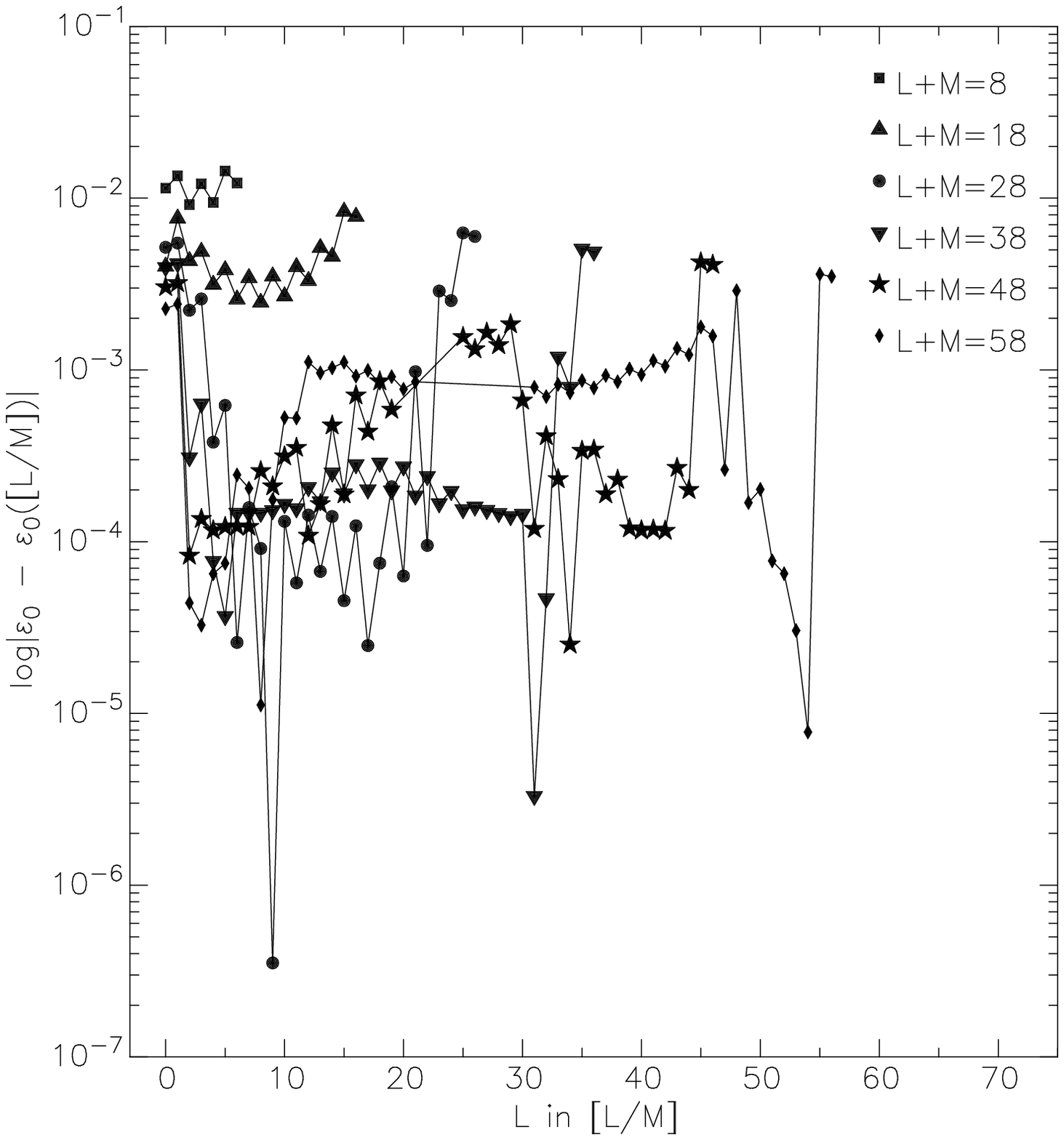}
 \caption
   [The error for the ground state energy computed using an Inversion
    analysis of $ dt(E)/dE $ with \hbox{$ [L/M] $} approximants, 
    versus $ L $ and $ L+M $ is fixed at various sizes.]{}
\end{figure}
\eject

\end{document}